\documentclass[MSNbibl,nameyear,dvips]{arxstspdf}
\usepackage{flushend}
\usepackage{stfloats}
\usepackage{graphicx}

\volume{28}
\issue{4}
\pubyear{2013}
\firstpage{641}
\lastpage{658}
\doi{10.1214/13-STS444} 

\makeatletter

\setattribute{abstract}    {skip} {18\p@}
\setattribute{keyword}     {skip} {6\p@}
\setattribute{abstract}   {width}  {370pt}
\setattribute{keyword}    {width}  {370pt}
\setattribute{frontmatter} {skip} {0\p@ plus 3\p@ minus 3\p@}

\newcommand{\citecs}[1]{\citeauthor{#1}, \citeyear{#1}}
\renewcommand{\citep}[1]{(\citeauthor{#1}, \citeyear{#1})}

\newcommand{\bfbeta}{\bolds{\beta}}
\newcommand{\btheta}{\bolds{\theta}}

\makeatother

\begin{document}
\begin{frontmatter}

\title{Scaling Integral Projection Models for Analyzing Size
Demography}
\runtitle{Scaling IPMs}

\begin{aug}
\author[a]{\fnms{Alan E.} \snm{Gelfand}\ead[label=e1]{alan@stat.duke.edu}},
\author[b]{\fnms{Souparno} \snm{Ghosh}\corref{}\ead[label=e2]{souparno.ghosh@ttu.edu}}
\and
\author[c]{\fnms{James S.} \snm{Clark}\ead[label=e3]{jimclark@duke.edu}}
\runauthor{A. E. Gelfand, S. Ghosh and J. S. Clark}

\affiliation{Duke University, Texas Tech University and Duke University}

\address[a]{Alan E. Gelfand is Professor,
Department of Statistical Science,
Duke University, Durham, North Carolina 27708, USA
\printead{e1}.}
\address[b]{Souparno Ghosh is Assistant Professor,
Department of Mathematics and
Statistics, Texas Tech University,
Lubbock, Texas 79409, USA \printead{e2}.}
\address[c]{James S. Clark is Professor, Nicholas School of the
Environment, Duke University,
Durham, North Carolina 27708, USA \printead{e3}.}
\end{aug}

%
\begin{abstract}
Historically, matrix projection models (MPMs) have been employed to
study population dynamics with regard to size, age or structure. To
work with continuous traits, in the past decade, integral projection
models (IPMs) have been proposed. Following the path for MPMs,
currently, IPMs are handled first with a fitting stage, then with a
projection stage. Model fitting has, so far, been done only with
individual-level transition data. These data are used in the fitting
stage to estimate the demographic functions (survival, growth,
fecundity) that comprise the kernel of the IPM specification. The
estimated kernel is then iterated from an initial trait distribution to
obtain what is interpreted as steady state population behavior. Such
projection results in inference that does not align with observed
temporal distributions. This might be expected; a model for population
level projection should be fitted with population level transitions.

Ghosh, Gelfand and Clark [\textit{J. Agric. Biol. Environ. Stat.}
\textbf{17} (2012) 641--699] offer a remedy by viewing the observed
size distribution at a given time as a point pattern over a bounded
interval, driven by an operating intensity. They propose a three-stage
hierarchical model. At the deepest level, demography is driven by an
unknown deterministic IPM. The operating intensities are allowed to
vary around this deterministic specification. Further uncertainty
arises in the realization of the point pattern given the operating
intensities. Such dynamic modeling, optimized by fitting data observed
over time, is better suited to projection.

Here, we address scaling of population IPM modeling, with the objective
of moving from projection at plot level to projection at the scale of
the eastern U.S. Such scaling is needed to capture climate effects,
which operate at a broader geographic scale, and therefore anticipated
demographic response to climate change at larger scales. We work with
the Forest Inventory Analysis (FIA) data set, the only data set
currently available to enable us to attempt such scaling.
Unfortunately, this data set has more than $80\%$ missingness; less
than $20\%$ of the 43,396 plots are inventoried each year. We provide
a hierarchical modeling approach which still enables us to implement
the desired scaling at annual resolution. We illustrate our methodology
with a simulation as well as with an analysis for two tree species, one
generalist, one specialist.
\end{abstract}

%

\end{frontmatter}

\section{Introduction}\label{sec1}

Population dynamics is a field with a long history in ecology and
biology. Demography summarizes traits classified as stages.
Matrix population models (MPMs), which assume stages are discrete
classes, are usually used to describe changing structure (see,
e.g., \citecs{Key85}, and references therein). Though
changes in trait operate at the individual level, analysis of
changing structure requires a translation of individual level data to
the population level.
In the past decade, the integral projection model (IPM)
({Easterling, Ellner and
Dixon}, \citeyear{EasEllDix00}; Ellner and Rees \citeyear{EllRee06},
\citeyear{EllRee07}; {Rees and Ellner}, \citeyear{ReeEll09}) has been offered as an
alternative to matrix projection models
when investigating continuous traits, for example, size, age, mass,
leaf length. These models are built from demographic functions,
parametric models for demographic processes specified in the form
of vital rates like growth, maturation, survival, birth and fecundity;
these rates are incorporated into a stationary
redistribution kernel. For an estimated demographic model, \emph
{projection} refers to iterative projection of this kernel to
steady state in order to attempt to answer questions regarding \emph
{what would happen}. So far, these models have only been
fitted with individual-level transition data, that is, these data are
used to estimate the demographic functions that comprise the
\emph{kernel} of the IPM specification. Then, projection proceeds
through iteration, given the estimated kernel.

In recent work, \citet{GhoGelCla12} argue that such an approach
introduces an inherent mismatch in scales.
Working with tree diameters as the trait of interest, an individual
level model describes the (conditional) transition of an
individual of diameter $x$ at time $t$ to diameter $y$ at time $t+1$.
On the other hand, an IPM essentially takes the
distribution of diameters of individuals at time $t$ to the
distribution of diameters of individuals at time $t+1$. We have a
version of the familiar ecological fallacy \citep{Wak09}. Moreover,
in our application, we do not have any individual level
transition data to attempt to scale up. For a given species, we only
have the collection of diameters in a given plot, in a
given year.

\citet{GhoGelCla12} offer a remedy by viewing the observed diameter
distribution at a given time as a point pattern over a
bounded interval, driven by an operating intensity. They propose a
three-stage hierarchical model. At the deepest level,
demography is driven by an unknown deterministic IPM. The operating
intensities vary around this deterministic specification.
Further uncertainty arises in the realization of the point pattern
given the operating intensities. \citet{GhoGelCla12} argue
that such dynamic modeling, optimized by fitting data observed over
time, will better reveal how intensities, hence population
structure, change over time. With individual-level IPM model fitting,
there is no mechanism to align projected trait
distributions with trait distributions observed over time;
consequential drift relative to the observed data can occur.


The contribution of this paper is to address scaling for population
level IPMs with the objective of moving from plot level scale
to larger scales, for example, the scale of the eastern U.S. Such
scaling is intended to try to identify climate effects, which operate
at a broader geographic scale, on demography. In turn, such scaling
could allow assessment of changes in trait distributions and
abundance, in response to climate change at larger scales. The threat
of climate change is typically evaluated in terms of
changes in distribution and abundance at regional scales (e.g., \citecs
{GuiRah11}). 
Our point pattern approach is attractive for scaling since we can
cumulate intensities to explain aggregated point patterns.
Individual level models cannot offer the desired scaling. Aggregation
has to be done in climate space since the IPM kernels
have arguments over trait space, not in geographic space, but introduce
climate as covariates.%

A further, critical contribution is an approach\break  to handle severe
sparsity in data collection. This\break  emerges as a key feature with
the USDA Forest Service's Forest Inventory and Analysis (FIA) data,
which motivates our scaling objective. The FIA data is one of
the few available data sets to investigate such scaling. Unfortunately,
this database has enormous missingness. Of the roughly
$44\mbox{,}000$ plots, less than $20\%$ are inventoried each year; more than
$80\%$ of plot level data over the time period 2000--2010 is
missing. During this period a plot will have been inventoried two,
possibly three times. Moreover, we have even more sparse
sampling prior to 2004.
Our study region includes the 31 eastern US states with climate
conditions varying from hot and moist near the Gulf of Mexico to
cold and dry near the Great Lakes. We focus on two illustrative
species: Acer rubrum (ACRU) is a generalist and occupies a wide
range of climate conditions. Liriodendron tulipifera (LITU) is
restricted to the hot moist climate of the eastern and
southeastern United States. Only in the recent work of
\citet{GhoGelCla12} have IPM models been considered at the population
scale. We are unaware of any applications at large geographic scales or
in the absence of data for consecutive time periods.

The format of the paper is as follows. In Section~\ref{sec2} we review
IPMs and their properties as well as how we introduce uncertainty into
the deterministic IPM specification. We also clarify the associated
model fitting challenge, offering an approximation. To expedite flow
and clarify the contribution here, explicit details of the model
specification, as developed in \citet{GhoGelCla12}, are deferred
to the \hyperref[app]{Appendix}. In Section~\ref{sec3} we describe the
FIA data set, as well as the climate data. In Section~\ref{sec4} we
develop the scaling model ideas, first with a full data set, then for a
very sparse data set. In Section~\ref{sec5} we provide a simulation
investigation to demonstrate the loss of information due to the severe
missingness as well as an analysis of the FIA data set. We offer some
concluding remarks in Section~\ref{sec6}.

\section{Integral Projection Models}\label{sec2}

In this section we briefly review the MPM, then turn to the IPM and its
properties. We discuss how to introduce uncertainty into
the IPM specification and conclude with a short discussion of how we
fit these models. Again, MPMs and IPMs are techniques of
choice for ecological demography. These models are specified with two
indices, one for time, the other for trait level, for example,
size, age, stage. There can be continuity or discreteness in time as
well as continuity or discreteness in the trait space. With
discrete time and categorical trait space, we have a MPM; with discrete
time and continuous trait space, an IPM. As Sections~\ref{sec2.1}
and \ref{sec2.2} reveal, MPMs and IPMs are not associated with a specified
region. There is no \emph{spatial} index in these models. This
reveals that scaling cannot be done over geographic space. A different
approach is needed, which we propose in Section~\ref{sec4}.

%

\subsection{Matrix Projection Models}\label{sec2.1}

MPMs specify population structure dynamically. The state of the
population at time $t$, as a vector of binned cell counts,
$\mathbf{n}(t)$, is multiplied by a population projection matrix,
$\mathbf{A}$, to yield the state of the population at time
$t+1$,
%
%
\begin{equation}\label{intrompm}\label{equ1}
\mathbf{n}(t+1)=\mathbf{A}\mathbf{n}(t).
\end{equation}
If the projection matrix is assumed to be time-invar\-iant, a linear
system of difference equations results to describe the
evolution of the population. When one allows the projection matrix to
vary because of external factors independent of the state
of the population, a more general, time-varying difference equation
version is obtained. If the projection matrix depends on the
current state of population itself, denoted by $\mathbf
{n}(t+1)=\mathbf{A}_n\mathbf{n}(t)$, we obtain a nonlinear model
termed a
density-dependent MPM. \citet{TulCas97} and \citet{Cas01}
discuss the features of all these MPMs in detail.
\citet{C08} examines change in response of nonlinear matrix
population models to changes in its parameters.

In (\ref{intrompm}), the $\mathbf{A}_{ij}$ give the average
per-capita contribution from individuals in category $j$ at time $t$
to category $i$ at time $t + 1$, either by survival, growth or
reproduction. Typically, $\mathbf{A}$ is written as
$\mathbf{A}=\mathbf{T}+\mathbf{F}$ with $\mathbf{T}$ describing
transition (survival and growth) and $\mathbf{F}$ describing
reproduction (fecundity). The stationary behavior of this matrix
projection equation is obtained in terms of the eigenvalues
($\Lambda_i$) and eigenvectors ($\mathbf{w}_i$) of $\mathbf{A}$. The
long-term (\emph{ergodic}) behavior of $\mathbf{n}(t)$ is
determined by the dominant eigenvalue, $\max(\Lambda_i)$, and
associated right eigenvector (see the book of \citecs{Cas01}, for
further details). Further eigenanalysis of the projection matrix yields
a set of population statistics, viz., population growth
rate, damping ratio, reproductive value and so on \citet{Cas01}. When
the model is density dependent, the resulting behavior of
the matrix equation cannot be written in terms of eigenvalues and
eigenvectors (\citecs{Cas01}, page 504). In fact, equilibrium
behavior need not exist.

\subsection{The IPM and its Properties}\label{sec2.2}

For continuous traits and, in particular, for diameters, the MPM
classes/stages are ordinal with definition being somewhat
arbitrary. In this regard, \citet{EasEllDix00} and \citet
{EllRee06} note that the IPM is proposed to remove the
categorization required under the MPM approach. Here, we briefly review
the behavior of an IPM as a deterministic specification.
Working with intensities, $\gamma_{t}(\cdot)$, subscripted by time,
we replace the MPM with
%
%
\begin{equation}\label{IPM1}\label{equ2}
\gamma_{t+1}(y) = \int_{L}^{U} K(y;x)
\gamma_{t}(x) \,dx.
\end{equation}
The kernel $K(y,x)$ is the IPM analog of the projection matrix $\mathbf
{A}$ in a MPM. $L$ and $U$ are the lower and upper limits
for the values of the trait.\footnote{Formally, \emph{finite} point
patterns are associated with bounded domains to ensure that
the integral of the intensity over the domain is finite. In one
dimension this means confining the support for $\gamma$ to a
bounded interval. Adopting this restriction in our setting implies that
$L$ can be 0, but we take $U < \infty$. This is a mild
practical constraint.} Note that $\gamma_{t}(x)$ is an intensity at
time $t$ implicitly associated with some region which we
will refer to as a \emph{plot}. So, if we integrate $\gamma_{t}(x)$
over an interval of diameters, we obtain the expected number
of individuals in the plot with diameters in that interval, at time
$t$. Therefore, $\gamma_{t,\cdot}= \int_{L}^{U} \gamma_{t}(x)\,dx$
is the expected number of individuals (population size) for the plot,
at time $t$. Integrating (\ref{IPM1}) over $y$ from $L$ to $U$
yields $\gamma_{t+1,\cdot} = \int_{L}^{U} K(\cdot,x) \gamma_{t}(x) \,dx$,
where $K(\cdot,x) = \int_{L}^{U} K(y;x) \,dy$; $\gamma_{t+1,\cdot}$
can be compared with $\gamma_{t,\cdot}$. To give a population level
interpretation to (\ref{IPM1}), it may be easiest to think in terms of
intensity elements. That is, $ \gamma_{t+1}(y) \,dy = \int_{L}^{U}
K(y;x) \,dy \gamma_{t}(x) \,dx$. But\break  then, we see that $K(y;x)
\gamma_{t}(x) \,dy \,dx$ is the expected number of individuals in diameter
interval $(y, y+dy)$ at time $t+1$ from all individuals in
diameter interval $(x, x+dx)$ at time $t$.

The eigenvalue theory for the IPM is directly connected to that for the
MPM by viewing $K(y;x)$ as a linear operator, that is, $Kh
\equiv\int_{L}^{U} K(y;x) h(x) \,dx$. If $\Lambda$ is the largest
eigenvalue associated with $K$ and $w(x)$ is the associated
right eigenfunction, $\int_{L}^{U} K(y;\break x)\cdot w(x) \,dx = \Lambda w(y)$,
showing that, at steady state, $\Lambda$ is the growth rate and
$w(x)$ (normalized) is the steady state diameter distribution.\footnote
{The Perron--Frobenius theory tells us that, at this $\Lambda$,
$w(x) \geq0$ $\forall x$.} As a result,\break  $K^{t}w =
\Lambda^{t}w$, but for arbitrary initial diameter distribution
$\gamma_{0}(x)$, the projection $K^{t}\gamma_{0}$ need not be close
to the projection $\Lambda^{t}\gamma_{0}$.


%
%
Specification of $K$ introduces a survival and growth term as well as a
fecundity or recruitment term. In practice, a
time-independent $K$ is not plausible; we employ a $K_{t}$ that is
dependent on levels of suitable environmental variables, say,
$\mathbf{Z}_{t}$, during year $t$, as well as density dependent, that
is, a function of $\gamma_{t,\cdot}$. Details on how we introduce
these features into $K_t$ are supplied in the \hyperref[app]{Appendix}.
The foregoing
scaling challenge is apart from the form of $K_{t}(y;x)$
so, in the sequel, we treat $K_{t}$ generically.

Sometimes normalization is introduced into the IPM. For instance, we
might replace $K(y;x)$ in (\ref{IPM1}) with, say, $
K(y;x)/K(\cdot;x)$, a\vadjust{\goodbreak} normalized version.\break  However, this removes the
interpretation of $\gamma_{t}(x)$ as an intensity since it
imposes $\gamma_{t,\cdot}$ constant over $t$. Normalizing $\gamma_{t}(x)$
to the density $\tilde{\gamma}_{t}(x)=
\gamma_{t}(x)/\gamma_{t,\cdot}$ is also unattractive since it now
normalizes the resulting $\gamma_{t+1}(y)$ by $\gamma_{t,\cdot}$ rather
than by $\gamma_{t+1,\cdot}$.
%
%

\subsection{Introducing Uncertainty}\label{sec2.3}

As specified, the foregoing IPM is deterministic, raising the question
of where and how to insert uncertainty. Within the
Bayesian framework, a natural choice is to make the parameters random.
However, a~broader concern involves uncertainty
associated with the form of $K$ itself. Insisting that the IPM model is
correct (even with ``best'' parameter estimates) is too
restrictive. Rather, we view the outcome of the IPM as a sequence of
intensities, $\gamma_{t}(y)$. Then, the operating
intensity, $\lambda_{t}(y)$ (i.e., the intensity that drives the
observed point pattern a time~$t$), is assumed to vary around
$\gamma_{t}(y)$. It is easier and more direct to specify uncertainty
through the $\lambda$'s than through the $K$'s; the latter
will prove computationally infeasible. With regard to the intensities,
we write $\lambda_{t}(x) =
\gamma_{t}(x)e^{\varepsilon_{t}(x)}$, where $\varepsilon_{t}$ is a
stationary Gaussian process over $[L,U]$ with covariance function
$\sigma_{\varepsilon}^{2}\rho(\cdot, \phi)$ and mean $0$.\footnote{It
might be more natural to set the mean equal to
$-\sigma_{\varepsilon}^{2}/2$ so that $E(\varepsilon_{t}(x))=1$. However,
mean $0$ only implies a scaling of $\lambda$ relative to
$\gamma$.} We have a log Gaussian Cox process (\cite{MW04}).

To allow for time-varying redistribution kernels, at least two
approaches emerge. The first assumes that a vector of parameters is
randomly chosen at each time point so that $K_{t}$ takes the form
$K(y,x;\break  \btheta(t))$. This strategy is employed (with individual-level
data) in, for example, \citet{ReeEll09}, where, under parametric
modeling for $K(y,x; \btheta)$, the posterior for $\btheta$ provides
draws for $\btheta(t)$. These draws may be interpreted as providing
temporal random effects rather than parameter uncertainty.\break  A~more
general \emph{regression} approach is to assume that $K$ is specified
as a fixed parametric function but involving time-varying climate
covariates and density dependence. Again, specific choices that we
employ are supplied in the \hyperref[app]{Appendix}. In any event, we
note that propagation of intensities through the $K_{t}$'s will not
yield explicit forms (even for stationary $K$'s). In fact, starting
from time~$0$, at time $t$ we will have a $t$-dimensional integration
for~$\gamma_{t}$.

Last, we do not apply the $K_{t}(y,x)$ to $\lambda_{t}(x)$. Rather, we
allow the IPM to provide dynamics in a deterministic
fashion for the $\gamma_{t}$'s, again viewing the $\lambda_{t}$'s
driving the point patterns as varying around their respective
$\gamma_{t}$'s. This specification suggests a pseudo-IPM
approximation, as discussed in Section~\ref{sec2.4}. As a result, the
$\lambda_{t}$'s are conditionally independent\break  given the $\{\gamma
_{t}\}$'s. At the highest level, we assume the point patterns,
the $\mathbf{x}_{t}$'s, are conditionally independent given their $\{
\lambda_{t}\}$'s with a nonhomogeneous Poisson process
likelihood given by
%
%
\begin{equation}\label{truelikelihood}\label{equ3}\quad
[\mathbf{x}_t|\lambda_t]\propto\Biggl[\exp\biggl(-
\int_{L}^{U} \lambda_t(x)\,dx \biggr)
\prod_{i=1}^{n_t}\lambda_t(x_{ti})
\Biggr].
\end{equation}
Hence, across $t$, the observed diameters are conditionally independent
given $\lambda_{t}(x)$, but are marginally dependent due
to the log Gaussian Cox process mod\-el for $\lambda_{t}(x)$.

Modeling is initiated with $\gamma_{0}$, a kernel intensity estimate
\citep{Dig83}. Hence, the full posterior is proportional to
%
%
\begin{eqnarray}\label{equ4}\quad
&& \prod_{t=1}^{T}\bigl[\mathbf{x}_{t}|
\lambda_{t}(x), x \in[L,U]\bigr]
\nonumber
\\
&&\quad{}\cdot\prod_{t=2}^{T}\bigl[\lambda_{t}(x), x
\in[L,U]|\gamma_{t}(x), x \in[L,U], \sigma^2, \phi\bigr]
\\
&&\quad{}\cdot\bigl[\bigl\{\gamma_{t}(\btheta,\gamma_0),
t=1,\ldots,T\bigr\}\bigr] \bigl[\sigma^2\bigr] [\phi]
[\btheta].\nonumber
\end{eqnarray}
In (\ref{equ4}), the bracketed term involving $\{\gamma_{t}\}$ is a degenerate
distribution. It is employed here to denote the
deterministic functional specification for the $\gamma_{t}$'s given
the IPM and $\btheta$.

\subsection{Model Fitting}\label{sec2.4}

We handle the stochastic integral in (\ref{equ3}) by discretization, as
described in the \hyperref[app]{Appendix}. However, the model described
in (\ref{equ4}),
using (\ref{equ2}), is computationally demanding to fit. The challenge arises
because (\ref{equ2}), with $K_{t}(y;x)$ as in (\ref{equ10}) in the \hyperref[app]{Appendix},
does not have a closed form solution; we need to resort to numerical
integration to create the sequence of $\{\gamma_{t}(x) \}$.
Moreover, the dimension of the numerical integrations increases as the
number of time epochs increases and consequently will
result in an explosion of summations over time. An MCMC scheme will be
computationally prohibitive because we will have to
perform these integrations iteration by iteration. Following
\citet{GhoGelCla12}, we propose an approximate ``pseudo'' IPM
approach, using adjacent pairs of years, that allows us to handle
general $K_{t}$.

\begin{figure}

\includegraphics{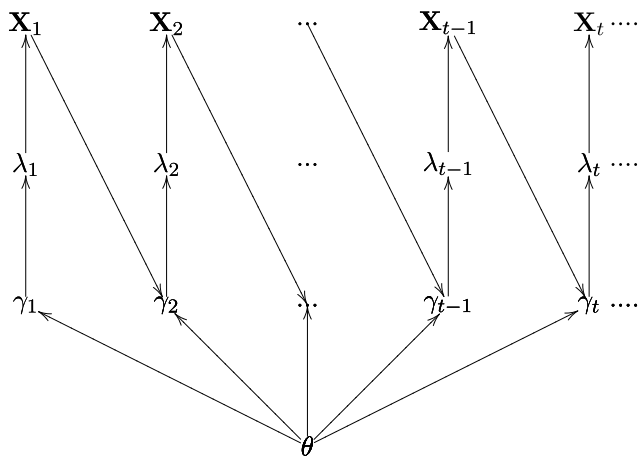}

\caption{Graphical model driving the dynamics in the ``pseudo''
IPM.}\label{fig1}
\end{figure}

As above (\ref{equ14}) in the \hyperref[app]{Appendix}, we consider $x^*_j$ to
be the center of\vadjust{\goodbreak}
the grid cell $j$. Then the pseudo-IPM update is given by
%
%
\begin{equation}\label{pseudoipm}\label{equ5}
\gamma_{t+1}\bigl(x^*_j\bigr) = \sum
_l K_t\bigl(x^*_j|x^*_l;
\mathbf{z}_t,\btheta,\gamma_{t,\cdot}\bigr)\hat{
\gamma_{t}}\bigl(x^*_l\bigr),
\end{equation}
where $\hat{\gamma_t}(x)$ is an empirical estimate of the intensity
corresponding to the point pattern observed at time point $t$
evaluated at the grid centers $x_l^*$. Under this updating scheme, for
each $t$, we replace the $t$-dimensional integral required
to get $\gamma_t(x)$ in (\ref{equ2}) by a one-dimensional integral. In (\ref{equ4}), to
obtain $[\{\gamma_{t}\};\btheta]$ would require computing
the $\gamma_{t}$ deterministically and sequentially for a given
$\btheta$, that is, $\prod_{1}^{T}[\gamma_{t}|\btheta, \gamma_{t-1}]$.
Using (\ref{equ5}), this now becomes\break  $\prod_{1}^{T}[\gamma_{t}|\btheta, \mathbf
{x}_{t-1}]$, where $\mathbf{x}_{t-1}$ yields
$\hat{\gamma}_{t-1}(x)$. The graphical model shown in Figure~\ref{fig1}
captures the pseudo-IPM approximation. By analogy,
%
%
pseudo-likelihood approximations in the literature often work with
pairs (in our case, years) of observations, often with good
asymptotic properties, though we cannot make such claims here.
However, the fact that our approximation from time $t$ to time
$t+1$ omits the uncertainty provided by $\lambda_{t}$ suggests that we
will underestimate uncertainty.

\section{Data Types and the FIA Data Set}\label{sec3}

In order to clarify the proposed scaling approach, we first describe
the motivating FIA data set as well as the associated climate
data. With MPMs, demographic data are customarily available at the
individual level over time. However, the case where the data
is in the form of a time series of population vectors is also discussed
(see \citecs{Cas01}).
With data of the latter type, we observe a sequence of population
vectors $\mathbf{n}(t_1),\mathbf{n}(t_2),\ldots$ without
distinguishing the individuals. Dennis et al. (\citeyear{Denetal95},
\citeyear{Denetal97}) use
nonlinear multivariate time-series methodology to obtain the
maximum likelihood estimates of the model parameters in this setting.
With IPMs, such data would consist of a time series of
point patterns for the trait distribution, for example, diameter, over
the study region.

For individual level data, perhaps the most direct modeling strategy
would be through a dynamic model with individual level
random effects as in, for example, \citet{Claetal10}. The state-space
framework provides inference on individual variation in terms of
population parameters, while being anchored directly by observations at
the same scale (or with specifications that translate
data to process scales, e.g., seed traps to trees). Afterward, desired
population-level summaries can be created.

%
%
\begin{figure*}
\begin{tabular}{@{}c@{\qquad}c@{}}

\includegraphics{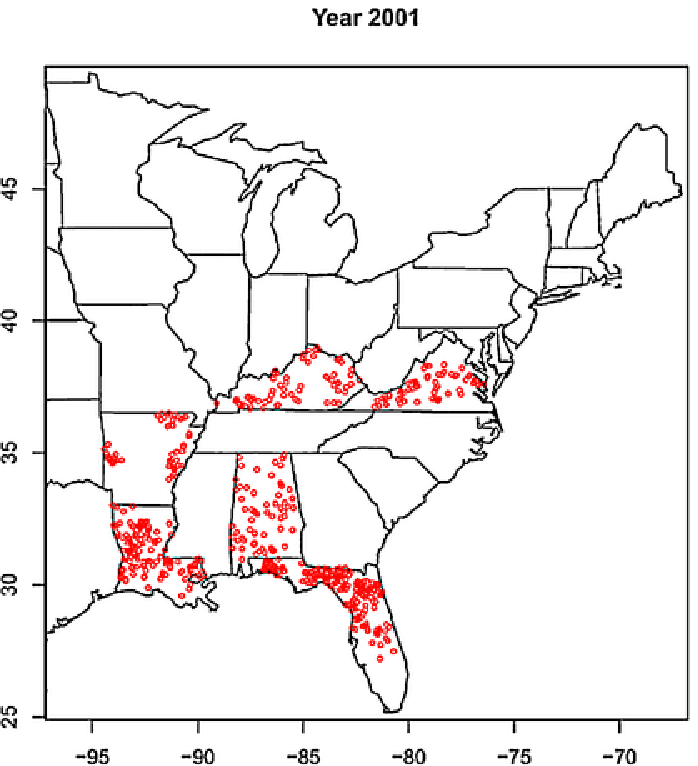}
 & \includegraphics{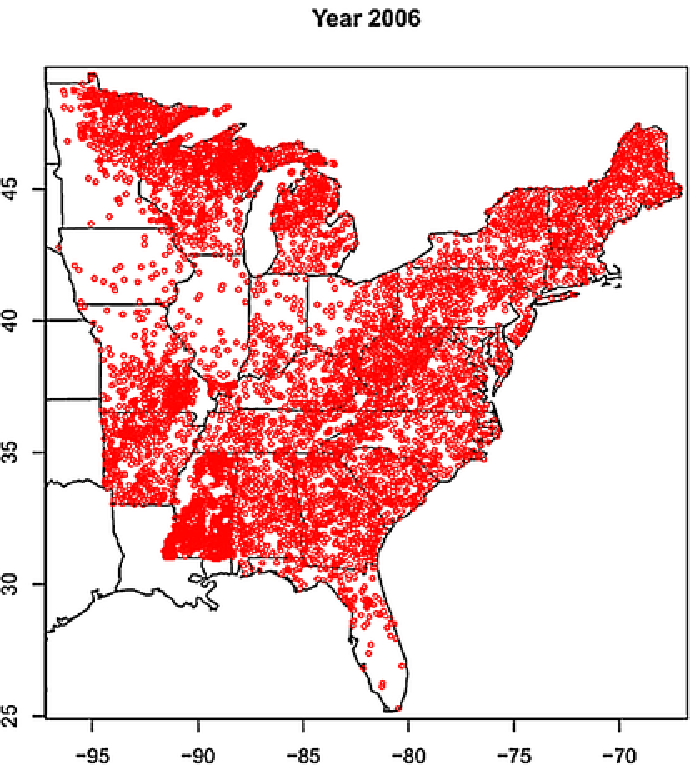}\\
(a) & (b)
\end{tabular}
\caption{Map of the sampled FIA plots in \textup{(a)} 2001 and \textup
{(b)} 2006.}\label{fig2}
\end{figure*}

However, at large geographic scales, tracking of individuals is not
feasible. For instance, we cannot hope to track individuals
in thousands of forests on an annual basis over a span of decades.
Collecting marginal point patterns at the scale of plots,
without transition information on individuals, is more realistic. Even
so, annual censusing of plots may not be. Hence, we need
to be able to fit IPMs with data of the this type.%

\subsection{The FIA Data}\label{sec3.1}

The USDA Forest Services Forest Inventory and Analysis (FIA) program is
the primary source for information about the extent,
condition, status and trends of forest resources in the United States
({Smith et~al.}, \citeyear{Smietal09}). FIA applies a nationally consistent
sampling protocol using a quasi-systematic design covering all
ownerships across the United States, resulting in a national
sample intensity of one plot per 2428 hectare \citep{BecPat05} where
plots are $54$ m$^2$. Aerial photography and/or
classified satellite imagery is used to stratify the population (i.e.,
increase the precision of population estimates) and to
establish permanent inventory plots in forest land uses. Forested land
is defined as areas at least 10\% covered by tree species,
at least 0.4 ha in diameter, and at least 36.6 m wide. FIA inventory
plots that are established in forested conditions consist of
four 7.2 m fixed radius subplots spaced 36.6 m apart in a triangular
arrangement with one subplot in the center \citep{BecPat05}. All trees
(standing live and dead) with a diameter at
breast height (dbh) of at least 12.7 cm are inventoried
on forested subplots. Within each subplot, a 2.07 m radius microplot
offset 3.66 m from subplot center is established where only
live trees with a dbh between 2.5 and 12.7 cm are inventoried. Within
each microplot, all live tree seedlings are tallied
according to species.

The program includes the measurement of a fixed proportion of the plots
in each state, in each year, known as annual inventory.
The legislative mandate requires measurement of 20\% of the plots in
each state, each year (FIA factsheet series, available
online,
\href{http://www.fia.fs.fed.us/library/fact-sheets/data-collections/Sampling\_and\_Plot\_Design.pdf}{http://www.fia.fs.fed.us/library/fact-sheets/}\break
\href{http://www.fia.fs.fed.us/library/fact-sheets/data-collections/Sampling\_and\_Plot\_Design.pdf}{data-collections/Sampling\_and\_Plot\_Design.pdf}).
In this analysis, the FIA
data we employ were extracted from the most recent annual inventories
(2000 to 2010) in 31 eastern states for a total of 43,396
inventory plots from FIADB version 4.0 on March 16, 2010 (available
online \url{http://fia.fs.fed.us/}).\break  The sampling is sparse in the
initial years. Collection increases starts in 2003 and reaches its
current level in 2006. The FIA plots sampled in year 2001
along with those sampled in 2006 are shown in Figure~\ref{fig2}. A
display of the set of plots sampled in 2007 would look almost the same
as in Figure~\ref{fig2}(b). Nonetheless, there would be no overlap
between the two sets of plots!

\subsection{The Climate Data}\label{sec3.2}

The climate data in this study was extracted from the 800m resolution
Parameter-elevation Regressions on Independent Slopes Model
(PRISM) data set\break  (available online
\href{http://www.prism.oregonstate.edu/}{http://www.prism.oregonstate.}\break
\href{http://www.prism.oregonstate.edu/}{edu/}). Recognized as the highest quality
spatial climate
data sets, PRISM is a sophisticated interpolation that uses
meteorological station data to produce continuous, digital grid
estimates of climatic parameters, with consideration of location,
elevation, coastal proximity, topographic facet orientation,
vertical atmospheric layer, topographic position and orographic
effectiveness of the terrain (\cite{Detal08}). In each
FIA-measured plot, we used the climate data from the previous year
to create the climatic covariates. We extracted the annual
average precipitation (in mm) and the mean winter temperature (in
$^\circ$C), the average of January, February and March maximum and
minimum monthly values.

Since climatological covariates operate over a\break  broad geographical area,
they may not explain the variation in, for example, diameter
distribution of a species at the plot level \citep{CanTho10}.
Such variation will likely depend more on micro-scale
covariates like soil moisture, nutrient\break  availability and so on, which
are not available to us. As a result, an approach to enable
climate to provide explanation of demography is through suitable
scaling. The scaling model described in the next section offers
a viable way to study diameter distribution with this objective. The
key idea is to partition climate space into bins.

\section{A Scaling Model}\label{sec4}

We first present the extension of the model in (\ref{equ2}) to enable the
desired scaling in the context of a full data set, that is, a
data set supplying annual diameter distributions (point patterns) for
every FIA plot. Then we present the model we employ to deal
with the extreme sparseness. A convenient way to appreciate the scaling
challenge is to envision a rectangular array where the
rows represent the plots, the columns denote the years, and, in a given
cell, we have a point pattern of diameters. In a full
data set, we have an observed point pattern in every cell. Imagining
this for the FIA data collection over ten years would entail
more than 400,000 point patterns. With the actual FIA data
collection, we have more than $80\%$ of the cells empty and for no
plot do we have point patterns in consecutive years.

\subsection{The Full Data Scaling Model}\label{sec4.1}

Again, we note that the IPM is indexed by time and by trait value but
there is no spatial index. In our context, there is
redistribution over trait space but no redistribution over geographic
space; the mod\-el specified in (\ref{equ2}) operates at the plot
level. With multiple plots, perhaps spatially-referenced, in principle,
we could fit this model plot by plot. While such an
analysis might be useful in some contexts, biogeographic studies
require joint modeling across plots. In other words, it would be
very difficult to develop a synthesis that would tell a big picture
demography story from such an analysis and, as noted above,
it would be difficult to capture climate effects. Moreover, it is
evidently not scaling the data to larger geographic regions.
In theory, with spatially-referenced plots, we might imagine
introducing plot-level spatial random effects into a joint model
across plots. However, with the FIA data, plots are not contiguous;
they are sparse across the eastern U.S. so such spatial
analysis is not appropriate.\footnote{An alternative might be to
introduce i.i.d. plot level random effects, but this will
substantially increase the dimension of the model and such a model be
difficult to fit, especially in our very sparse sampling
setting.} Rather, as noted earlier, an attractive feature of working
with point patterns and associated intensities in a Cox
process setting is the convenience of aggregating intensities to
explain aggregated point patterns. First, we consider how we
might do this with a full data set.

Recall that each plot is subjected to a sequence of annual climate
variables across the years of data collection. Suppose we
partition the climate space into a collection of climate bins, indexed
by $l=1,2,\ldots,M$. In the present setting, we have two
climate covariates and we are partitioning the upper right quadrant of
$\mathbb{R}^2$. (Below, we say more about the choice of
partition.) Suppose we label each climate bin by a suitable centroid
(defined below). Then, we have a set of $M$ climates which
are labeled as $\mathbf{z}_{l}^{*}$. Suppose, in year $t$, we assign
label $L_{j,t}=l$ to plot $j$ if it received climate
falling in bin $l$. Thus, in year $t$, all of these plots will have
their diameter distributions operated on by the same
redistribution kernel, that is, $K$ as in (\ref{equ10}) in the \hyperref
[app]{Appendix} with
$\mathbf{z}_{t} = \mathbf{z}_{l}^{*}$, apart from plot-specific
density dependence. In particular, suppose $S_{t,l} = \{j\dvtx
L_{j,t}=l\}$
and there are $n_{t,l}$ plots receiving climate
$\mathbf{z}_{l}^{*}$ in year $t$. Then, $\sum_{j \in S_{t,l}} \gamma
_{j,t}(x)$ is the cumulated intensity subjected to $K$, with
$\mathbf{z}_{t} = \mathbf{z}_{l}^{*}$ in year $t$. The only
modification is that, if density dependence appears in $K$, then the
population size for density dependence would be $\sum_{j \in S_{t,l}}
\gamma_{j,t,\cdot}$. We acknowledge that density dependence
operates at the plot scale, not at the scale of aggregation to climate
bins. However, at larger scale, it is still arguable that
an increase in aggregated abundance at year $t$ will place a scaled
increase in resource pressure on the species for year $t+1$.
Moreover, if we wish, we can compare models with and without aggregated
density dependence.

If we do the above for each $l$, then in year $t$, every plot will have
been assigned to a unique climate bin and we can fit the
IPM across the $M$ bins for year $t$. Then, if we do this for each
year, we have jointly fitted the IPM across all plots for all
years.

Finally, we assign as the ``centroid'' to bin $l$ the average of all of
the climates for all of the plots across all of the time
points that fell into bin $l$. That is, we keep the partitions the same
from year to year and we keep the labels constant across
time as well. As for the creation of the partitions, we overlay a
bounding rectangle on the observed climates for all plots and
all years in the study. We then partition the temperature axis (mean
winter temperature) as well as the precipitation axis
(average annual precipitation) to create a rectangular grid. 
Some bins will be empty in one or more years; they are not considered.
In fact, if, for a given species, there were no
occurrences across all plots in a climate bin in a given year, then the
bin is not considered for that year. We have no point
pattern to drive the pseudo-IPM for the species in that year.\vadjust{\goodbreak}

Again, we emphasize that we are scaling in climate space, not in
geographic space; we are aggregating plots receiving essentially
a common climate in a given year regardless of where they are in
physical space. But, this raises the question of what
projection means under such scaling? In climate space, we fit data
across $M \times T$ cells but it makes no sense to project a
climate bin in time. If we aggregated plots in geographic space, we
would not necessarily find common climate for all plots in
each year. So, we cannot do projection for such spatially aggregated
plots? The conclusion is that we should think of
projection at the plot level or for an aggregated collection of plots
receiving the same sequence of climate variables over time.
Projection under such aggregation may be adequate to address large
scale response to broad, coarse spatial resolution climate
scenarios. In any event, these limitations are not a criticism of our
approach to scaling. Rather, they clarify what projection
can entail. Moreover, as we shall see in the next subsection, our
approach offers the only way to implement demographic modeling
for the FIA data with its inherent sparseness.

To conclude here, we consider inference summaries under our modeling.
First, we can present posterior summaries of the model
parameters, that is, all of the parameters in the $K_{t}$'s. Next, we
can develop posterior predictive intensities to compare with
empirical intensities for different $(l,t)$ combinations. Also, we can
provide comparison of observed and predicted population
sizes under various $(l,t)$ combinations. Last, at the plot level, we
can implement projection under the model and compare
predicted intensities and population sizes with their observed counterparts.

\subsection{Accommodating the Sparseness in the FIA~Data}\label{sec4.2}

With regard to the discussion of the previous subsection, now, for
every species, our plot by year array has more then $80\%$ cells with
no observed point pattern. As above, for plot $j$ at time $t$, if
$\mathbf{z}_{j,t}$ is the associated covariate, we will assign label
$L_{j,t} = l$ if $\mathbf{z}_{j,t}$ falls in bin~$l$. So, if
$L_{j,t}=l$, the redistribution kernel operating at time $t$ is
$K_{t}(y,x; \mathbf{z}_{l},\boldsymbol{\theta})$.

We formalize our modeling at the plot level, that is, for plot $j$ in
year $t$, if $L_{j,t}=l$,
%
%
\begin{equation}\label{equ6}
\gamma_{j,t+1}(y) = \int K(y,x;\mathbf{z}_{l}, \boldsymbol{
\theta}) \gamma_{j,t}(x) \,dx.
\end{equation}
With the foregoing notation, for year $t$, we have $S_{t,l} = \{j\dvtx
L_{j,t}=l\}, l=1,2,\ldots,M$, with $n_{t,l}$, the number of plots
in $S_{t,l}$. The number of\vadjust{\goodbreak} plots in $\bigcup_{l} S_{t,l}$ is $
n_{t}$, the number of FIA plots in year $t$. (There is year to
year variation.) Next, let $I_{j,t} =1,0$ if plot $j$ is measured
(observed) or not in year $t$. That is, if $I_{j,t}=1$, we
observe a point pattern, $\mathbf{x}_{l,t}$. Then, let $S_{t,l,1} = \{
j\dvtx L_{j,t}=l, I_{j,t}=1\}$, $S_{t,l,0} = \{j\dvtx L_{j,t}=l,
I_{j,t}=0\}$. Evidently, $S_{t,l,1}\cup S_{t,l,0} = S_{t,l}$.
Similarly, define $R_{t,l,1} = \{j\dvtx L_{j,t}=l, I_{j,t+1}=1\}$,
$R_{t,l,0} = \{j\dvtx L_{j,t}=l,\break  I_{j,t+1}=0\}$. So, $R_{t,l,1}\cup
R_{t,l,0} = S_{t,l}$. The idea here is, for plots that
experienced $\mathbf{z}_{l}$ in year $t$, we wish to capture the set
which was observed at the \emph{start} of the year
($S_{t,l,1}$) and the set which was observed at the \emph{end} of the
year ($R_{t,l,1}$). Again, $S_{t,l,1}$ and $R_{t,l,1}$
are disjoint, providing the crux of the missing data challenge. Let
$n_{t,l,1}$ and $m_{t,l,1}$ denote the number of plots in
$S_{t,l,1}$ and $R_{t,l,1}$, respectively.

From (\ref{equ6}), we have the following conceptual IPM scaling:
%
%
\begin{eqnarray}\label{equ7}
&&
\sum_{j \in S_{t,l}} \gamma_{j,t+1}(y) \nonumber\\[-8pt]\\[-8pt]
&&\quad= \int
_{D} K(y,x; \mathbf{z}_{l},\boldsymbol{\theta})
\sum_{j \in S_{t,l}} \gamma_{j,t}(x) \,dx.\nonumber
\end{eqnarray}
Again, we only see a subset of the plots on the left-hand side of (\ref{equ7}) and
also only a subset of the plots on the right-hand side. However,
dividing both sides by~$n_{t,l}$, we have the ``per plot'' (or average)
intensity,
%
%
\begin{equation}\label{equ8}
\bar{\gamma}_{l,t+1}(y) = \int_{D} K(y,x; \mathbf
{z}_{l},\boldsymbol{\theta}) \bar{\gamma}_{l,t}(x) \,dx
\end{equation}
with the obvious definition for the $\bar{\gamma}$'s.

But, this leads to the natural approximations:\break  $ \tilde{\gamma
}_{l,t+1}(y) \approx\bar{\gamma}_{l,t+1}(y)$ and
$\tilde{\gamma}_{l,t}(x) \approx\bar{\gamma}_{l,t}(x)$, where\break
$\tilde{\gamma}_{l,t+1}(y) = \sum_{j \in R_{t,l,1}}
\gamma_{j,t+1}(y)/m_{t,l,1}$ and $\tilde{\gamma}_{l,t}(x) =\break  \sum_{j
\in S_{t,l,1}} \gamma_{j,t}(x)/n_{t,l,1}$.\vspace*{2pt}

So, for each $t$ and $l$, we work with the approximate IPM relationship,
%
%
\begin{equation}\label{scaledgamma}\label{equ9}
\tilde{\gamma}_{l,t+1}(y) = \int_{D} K(y,x; \mathbf
{z}_{l},\boldsymbol{\theta}) \tilde{\gamma}_{l,t}(x) \,dx.
\end{equation}
To use this relationship, we have to do two things:

\mbox{}\hphantom{i}(i) create an empirical estimate of $\tilde{\gamma}_{l,t}(x)$ with
the observed $\mathbf{x}_{j,t}$ for $j \in S_{l,t,1}$ to use
on the right-hand side. We do this by creating an empirical intensity based
upon all of the plots in $S_{t,l,1}$ and then scaling the
intensity by $n_{t,l,1}$.

(ii) use the observed $\mathbf{x}_{j,t+1}$ for $j \in R_{t,l,1}$ to
inform about the left-hand side. That is, the observed
$\mathbf{x}_{j,t+1}$ are linked to $m_{t,l,1} \lambda_{l,t+1}(y)$, as
above in expression (\ref{equ9}), in the likelihood and
$\lambda_{l,t+1}(y)$ is linked to $\tilde{\gamma}_{l,t+1}(y)$, up to
log GP error, as in Section~\ref{sec2.3}. To do this, we introduce a
``per plot'' log GP error for each climate bin in each year.

A further complication arises due to lack of information about new
recruits [again, see (\ref{equ10})--(\ref{equ12}) and related discussion in the
\hyperref[app]{Appendix}]. In the absence of consecutive years of data,
we cannot
distinguish if an individual observed at time $t+1$ in bin $l$
is actually a new recruit (crossing the boundary from seedling to
trees) or was in an unsampled plot in that bin at time $t$. Due
to this ambiguity, $\delta_1$ in (\ref{equ12}) cannot be estimated reliably.
Hence, the effect of density dependence on recruitment rate
cannot be ascertained at this level of sparsity. As a result, we set
$\delta_1=0$ and thereby assume a time-invariant $\Delta$.

As with a full data set but even more so, some blocks will be empty in
one or more years; they are not considered. Again, if, for
a given species, there were no occurrences across all plots in a bin in
a given year, then we have no point pattern to drive the
pseudo-IPM for that bin for that year. For that species, the bin is not
considered for that year. Inference summaries will
parallel those we proposed above to create with a full data set.

\section{A Simulation and an FIA Data Analysis}\label{sec5}

In Section~\ref{sec5.1} we provide a simulation to show how well our approach
works with a full data set and also to reveal the effect of
the loss of information as we go to $50\%$ and also $80\%$ missingness.
Then, in Section~\ref{sec5.2}, we turn to the FIA data, to look
at two species, recognizing the inference challenges imposed by the sparsity.

\subsection{A Simulation Example}\label{sec5.1}

To illustrate the performance of the model under various level of
missingness, we performed the following simulation study. We
envisioned a covariate with four levels assigned as $z = 0, 1, 2, 3$
and to each covariate level we assigned 100 FIA plots from
year 2005. Thus, the covariate level can be viewed as a climate,
defining four ``climate bins'' with, initially, 100 plots in
each bin. We used the empirical intensity associated with each plot
(from the 2005 FIA data), say, $\hat{\gamma}_{lj,0} (x^*),
l=1,2,3,4; j=1,2,\ldots,100$, as the initial condition, for
$x^{*}$'s as described in Section~\ref{sec2.4}. We plug this initial
intensity in the pseudo-IPM (\ref{pseudoipm}) with covariate
information inserted in the redistribution kernel (\ref{genK}) to
obtain $\gamma_{1}(\cdot)$ (for each plot) at the following time point.
With regard to $K$, we used the forms in the \hyperref[app]{Appendix}, fixing
$Q_{0}=1$, $\delta_{1}=0$, and $\mu=0$, with the remaining parameter
values set as in Table~\ref{tabl1}. We seek to infer about these
remaining parameters, including the regression coefficient $\beta$ for
climate, as well as to project 10 years forward. To mimic
the FIA data set, we randomize the plots to changing climate at each
time point following the illustrative transition matrix shown
below:\vspace*{9pt}
\begin{center}
{\fontsize{9pt}{10pt}\selectfont{
\begin{tabular}{@{}l l l l l@{}}\hline
& \multicolumn{4}{l}{Location at $t+1$}  \\ \hline
Location at $t$ & 0 & 1 & 2 & 3 \\
\hline
0 & 0.7 & 0.2 & 0.07 & 0.03 \\
1 & 0.2 & 0.7 & 0.03 & 0.07 \\
2 & 0.07 & 0.03 & 0.7 & 0.2 \\
3 & 0.03 & 0.07 & 0.2 & 0.7\\
\hline
\end{tabular}}}\vspace*{9pt}
\end{center}
In this fashion, for each plot, we generate a sequence of $\gamma
_t(x^*)$ for 10 time points and then an associated point
pattern. We then fit the scaling model (\ref{scaledgamma}) using the
first nine time points of the complete data set. The average
empirical intensity at each covariate bin at $t=1$ [$\tilde{\gamma
}_{l,1}(x^*)$] is used in fitting this full data model. The
posterior summary of the parameters (Table~\ref{tabl1}) suggests that,
when there
is no missingness, the scaling model can recover the
parameters of the plot level model.

%
%
\begin{table}
\caption{Posterior summaries of model parameters under various
levels of missingness for the simulated data. Posterior mean and $95\%$
equal tail credible intervals provided} \label{tabl1}
\begin{tabular*}{\tablewidth}{@{\extracolsep{\fill}}l ccl@{}}
\hline
& \textbf{True} & \textbf{Level of} & \\
\textbf{Parameters} & \textbf{value} & \textbf{missingness} &
\multicolumn{1}{c@{}}{\textbf{Posterior summary}}\\
\hline
$Q_{1}$ & 0.01 & \hphantom{0}0\% & 0.0171 (0.0058, 0.0292) \\
$\sigma_{}$ & 0.25 & & 0.2456 (0.1054, 0.3923)\\
$\delta_{0}$ & 0.30 & & 0.3467 (0.2161, 0.4860) \\
$\eta_{}$ & 0.10 & & 0.1253 (0.0548, 0.1953) \\
$ \beta$ & 0.01 & & 0.0091 (0.0034, 0.0148)\\
[4pt]
$Q_{1}$ & 0.01 & 50\% & 0.0174 (0.0016, 0.0275) \\
$\sigma_{}$ & 0.25 & & 0.2695 (0.0596, 0.4849)\\
$\delta_{0}$ & 0.30 & & 0.3281 (0.1181, 0.5643) \\
$\eta_{}$ & 0.10 & & 0.1242 (0.0530, 0.1962) \\
$\beta$ & 0.01 & & 0.0146 ($-$0.0002, 0.0292)\\
[4pt]
$Q_{1}$ & 0.01 & 80\% & 0.0293 (0.0125, 0.0582) \\
$\sigma_{}$ & 0.25 & & 0.5702 (0.2756, 0.8323)\\
$\delta_{0}$ & 0.30 & & 0.3536 (0.1783, 0.5315) \\
$\eta_{}$ & 0.10 & & 0.0824 (0.0089, 0.2201) \\
$\beta$ & 0.01 & & 0.0691 (0.0220, 0.1166) \\
\hline
\end{tabular*}
\end{table}

%
%
\begin{figure*}

\includegraphics{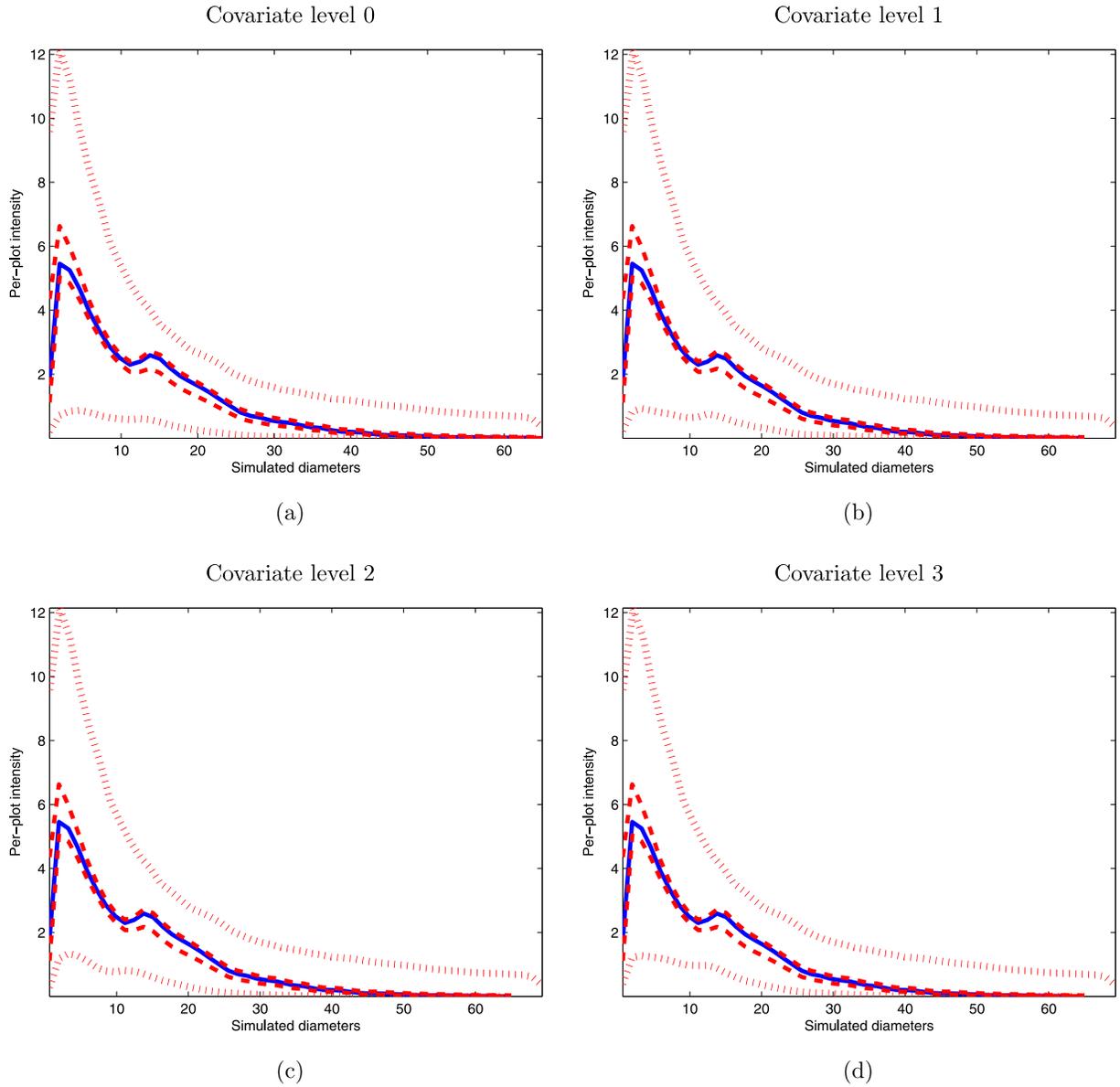}

\caption{Plot of true simulated $\tilde{\gamma}_{l,10}(x^*)$ (solid)
for four covariate bins (see text for details). Overlaid are the
pointwise 95\% CI of the projected $\gamma_{l,10}(x^*)$ under the
complete data set (dashed) and those under the sparse data set with
80\% missingness per time point (dotted). The covariate levels are
noted atop the figures. Note: the posterior medians are not displayed
in Figure~\protect\ref{fig3}. Those figures are available upon
request.}\label{fig3}\vspace*{6pt}
\end{figure*}

In order to investigate our ability to recover the parameters under a
\emph{moderate} amount of missingness, we randomly remove
50 plots from each covariate bin at each of the first nine time points
and fit the scaling model (\ref{scaledgamma}) using the
remaining available 200 plots at each time point in the training set.
The posterior summary in Table~\ref{tabl1} shows that we can still
recover certain parameters, although, as expected, the uncertainty
associated with these estimates is higher than that obtained
for the complete data set. Turning to \emph{extreme} missingness, as
in the FIA data, we randomly remove 80 plots from each
covariate bin at every time point in the training set and then fit the
scaling model to the remaining plots. The posterior
summary in Table~\ref{tabl1} shows that the intervals are now even
longer and not
well centered. In summary, our modeling approach is
viable but, with very high levels of missingness, our ability to learn
about the process will be limited.

Next, we illustrate how missingness leads to increased uncertainty in
projection. Let $\Theta^{\mathrm{comp}}$ and $\Theta^{\mathrm{ext}}$ be
posterior samples of all the parameters obtained from fitting the
scaling model on the complete data set and the data set with 80\%
missing plots, respectively. Using $\Theta^{\mathrm{comp}}_b$, the $b$th
posterior sample of $\Theta^{\mathrm{comp}}$ and
$\tilde{\gamma}_{l,1}(x^*)$, we generate $\tilde{\gamma
}_{l,2}(x^*;\break \Theta^{\mathrm{comp}}_b),\ldots,
\tilde{\gamma}_{l,10}(x^*,\allowbreak\Theta^{\mathrm{comp}}_b), l=1,2,3,4;
b=1, 2,\break\ldots,B$. Similarly, using $\Theta^{\mathrm{ext}}_b$, we generate
$\tilde{\gamma}_{l,1}(x^*;\Theta^{\mathrm{ext}}_b),\break \ldots, \tilde{\gamma
}_{l,10}(x^*,\Theta^{\mathrm{ext}}_b)$. Figure~\ref{fig3} shows the true
$\tilde{\gamma}_{l,10}(x^*)$ along with the pointwise 95\% CI
obtained for\break  $\tilde{\gamma}_{l,10}(x^*,\Theta^{\mathrm{comp}}_b)$ and
$\tilde{\gamma}_{l,10}(x^*,\Theta^{\mathrm{ext}}_b)$ for $l=1,2,3,4$. In all
cases we are able to contain the true ten year projections
under the extreme missingness but, as expected, the uncertainties
associated with the projection using $\Theta^{\mathrm{ext}}$ are
substantially higher than those obtained for $\Theta^{\mathrm{comp}}$.

\subsection{Data Analysis for Two Species}\label{sec5.2}

We illustrate the scaling model on two species \textit{Acer rubrum} (ACRU)
and \textit{Liriodendron tulipifera} (LITU). ACRU has a
broad geographic distribution, its range extending from the Gulf Coast
of the eastern United States to Canada. It can thrive on
mesic (moderate moisture) to xeric (dry) sites. Compared with ACRU,
LITU is much less widespread. Its range does not extend as
far north, nor does it occupy xeric sites.

\begin{figure*}

\includegraphics{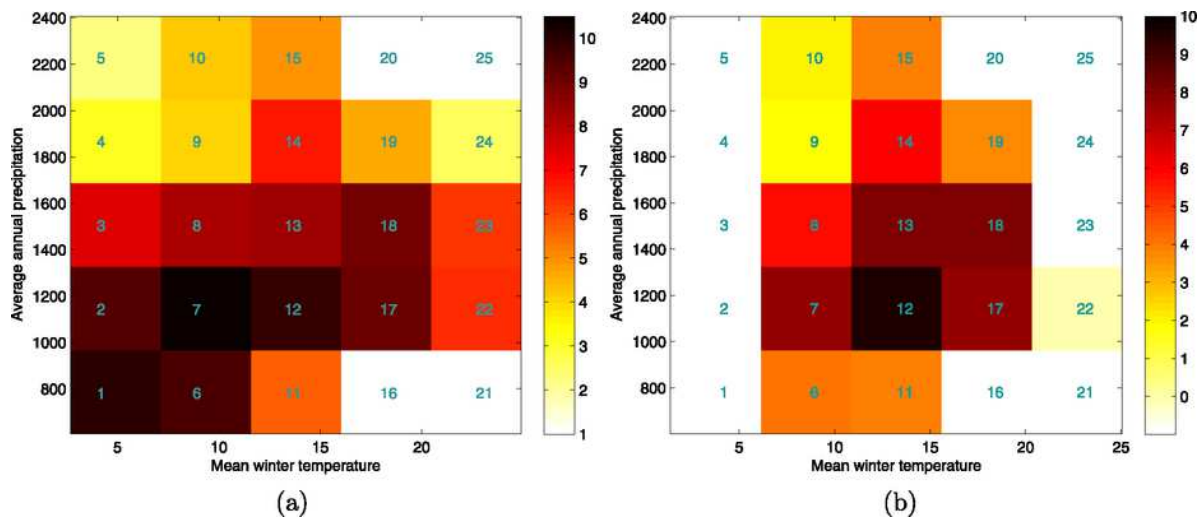}

\caption{The climate bins along with the logarithm of the number of
plots in which \textup{(a)} ACRU were observed during the period of
study (from 2005 to 2010) in each bin, and \textup{(b)} LITU were observed
during the period of study (from 2005 to 2010) in each bin. Also,
the bins are indexed from $1$ to $25$.}\label{fig4}
\end{figure*}

Due to the extreme sparsity of sampling prior to 2005, the scaling
model is fitted on data from 2005 through 2010. We use the
data from 2005 to provide the initial intensities. We partition the
available information (from the PRISM data set) on pairs of
mean winter temperature and average annual precipitation for the period
under study in a $5\times5$ equi-spaced grid and use
the ``centroid'' (see Section~\ref{sec5.1}) of each bin as~$\mathbf{z}_l$.
Figures~\ref{fig4}(a) and \ref{fig4}(b) show climate bins for ACRU and LITU,
%
%
respectively, along with the logarithm of the total number of plots
observed in each bin, for each species during the period of
study. We see that, for ACRU and, even more so, for LITU, there are
climate bins in which the species were not observed. Also, for
convenience in display, the bins are numbered from $1$ to $25$ as indicated.

For display purposes, we choose four climate bins that have at least
100 individuals in each of the years under study. Figure~\ref{fig5a}
shows the summary of the estimated per-plot intensities for ACRU for
%
%
\renewcommand{\thefigure}{5\textup{a}}
\begin{figure*}

\includegraphics{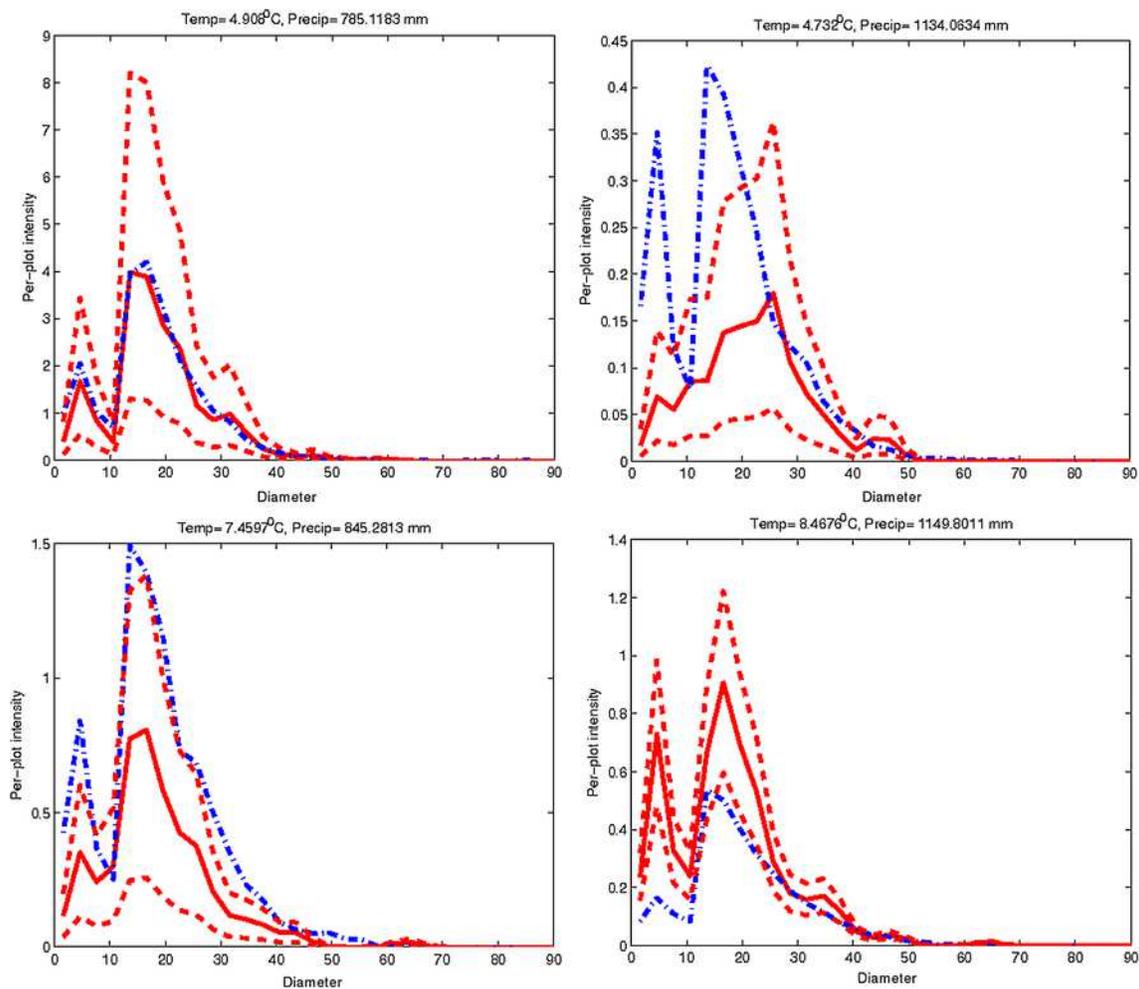}

\caption{Posterior mean (solid) and pointwise 95\% CI (dashed) for
the estimated per-plot intensity for ACRU for 2006. The
observed empirical per-plot intensity is overlaid (dash-dotted). The
temperature and precipitation corresponding to the grid
centroid are noted atop.}\label{fig5a}
\end{figure*}
these grids for 2006. The corresponding observed empirical
per-plot intensity is overlaid. Figure~\ref{fig5b} shows the same for
year 2009.
Figures~\ref{fig6a} and \ref{fig6b} are similar to Figures~\ref{fig5a}
and \ref{fig5b} but
for the species LITU. Two remarks emerge, neither unexpected in view of
the severe sparsity. We evidently do better in some
climate bins than others and we have a very large amount of
uncertainty.

%
%
\renewcommand{\thefigure}{5\textup{b}}
\begin{figure*}

\includegraphics{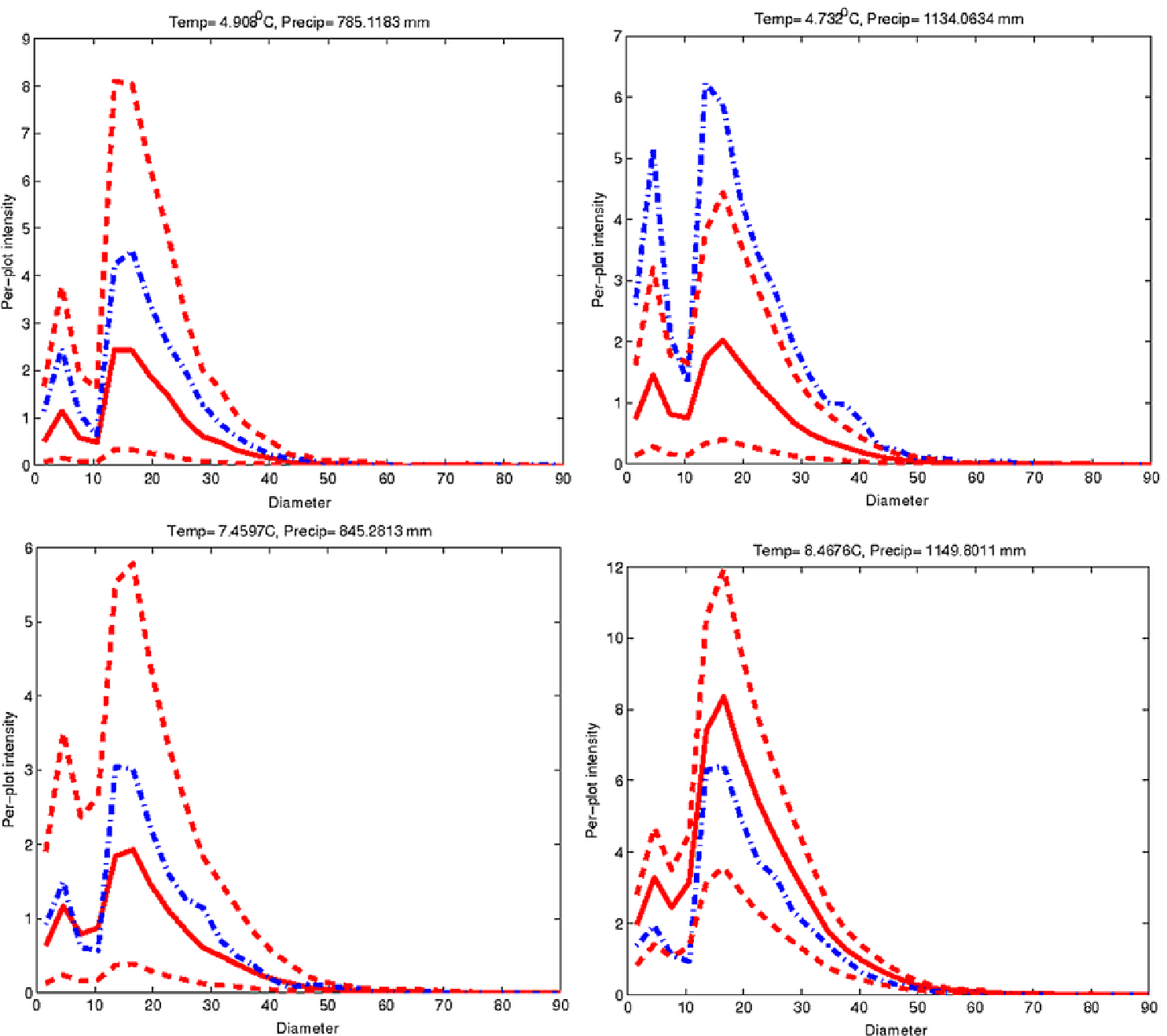}

\caption{Posterior mean (solid) and pointwise 95\% CI (dashed) for
the estimated per-plot intensity for ACRU for 2009. The
observed empirical per-plot intensity is overlaid (dash-dotted). The
temperature and precipitation corresponding to the grid
centroid are noted atop.}\label{fig5b}
\end{figure*}

%
%
\renewcommand{\thefigure}{6\textup{a}}
\begin{figure*}

\includegraphics{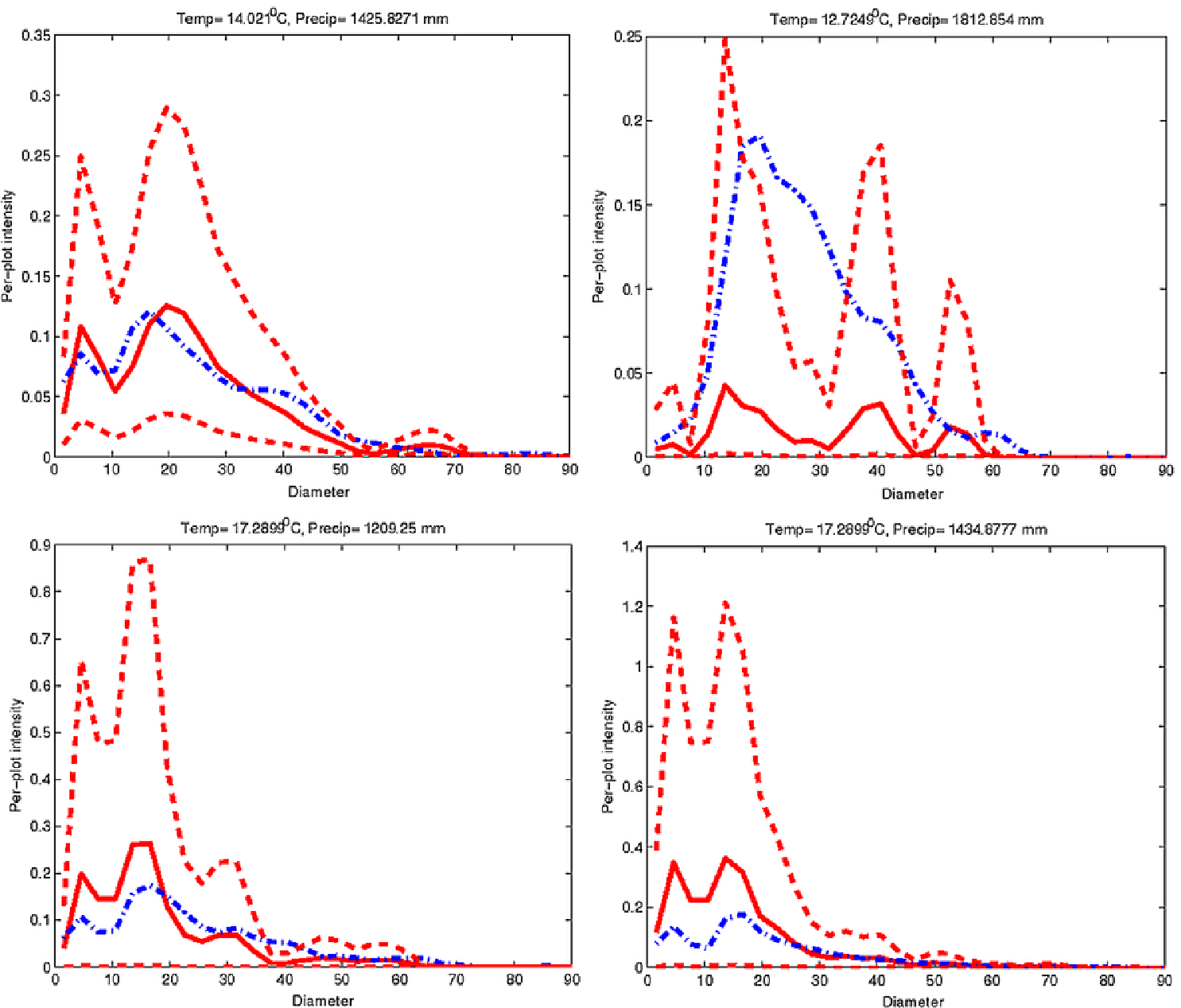}

\caption{Posterior mean (solid) and pointwise 95\% CI (dashed) for
the estimated per-plot intensity for LITU for 2006. The
observed empirical per-plot intensity is overlaid (dash-dotted). The
temperature and precipitation corresponding to the grid
centroid are noted atop.}\label{fig6a}\vspace*{6pt}
\end{figure*}

%
%
\renewcommand{\thefigure}{6\textup{b}}
\begin{figure*}

\includegraphics{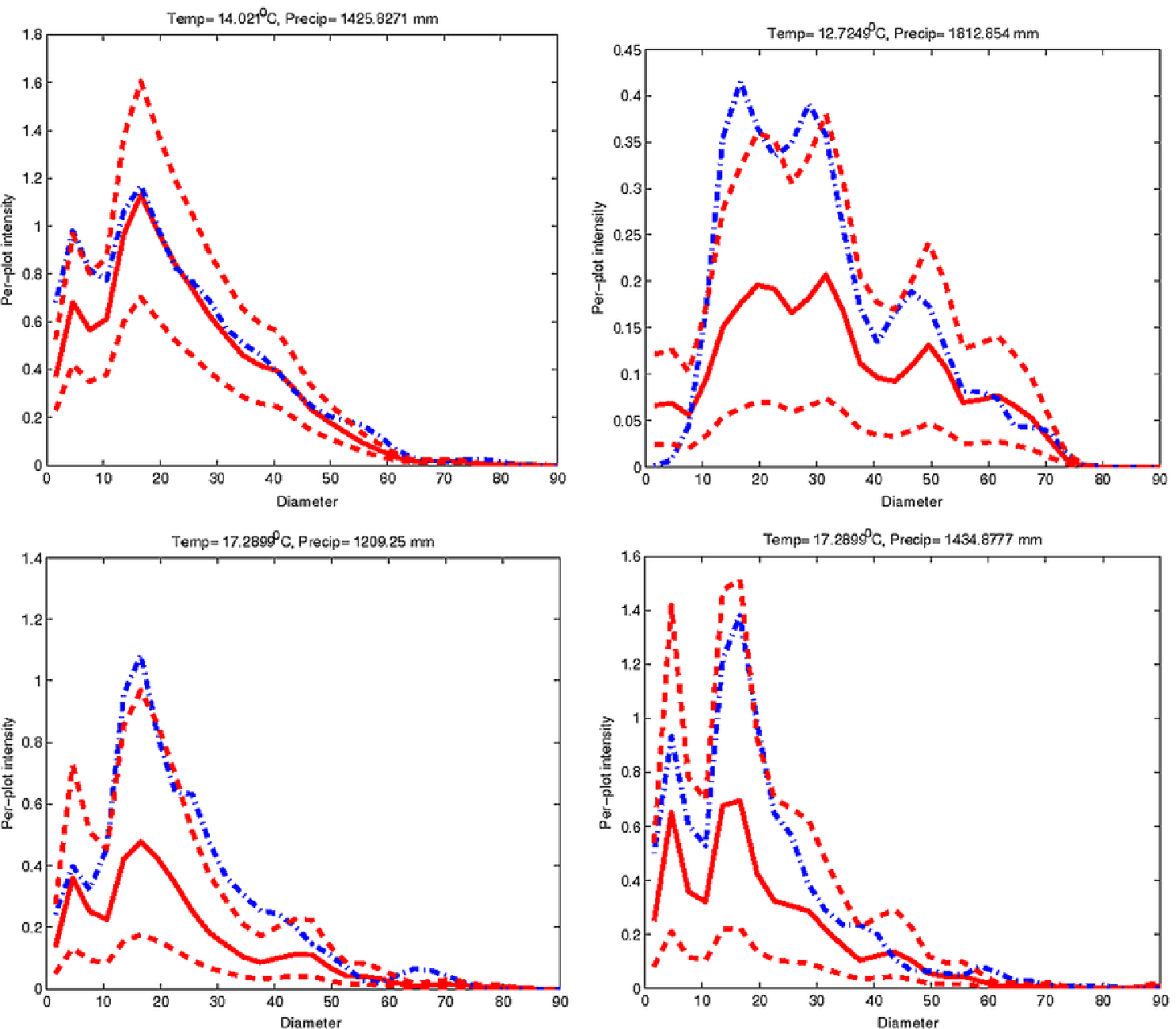}

\caption{Posterior mean (solid) and pointwise 95\% CI (dashed) for
the estimated per-plot intensity for LITU for 2009. The
observed empirical per-plot intensity is overlaid (dash-dotted). The
temperature and precipitation corresponding to the bin
centroid are noted atop.}\label{fig6b}
\end{figure*}

To further assess the goodness of fit, we plot the posterior summary of
the per-plot estimated population size for nonempty
climate bins (indexed as in Figure~\ref{fig4}) in Figures~\ref{fig7a}
and \ref{fig7b}. Overlaid
are the observed per-plot population size in the
corresponding climate grid. Generally, our prediction is successful
but, again, our uncertainty is very high.

%
%
\renewcommand{\thefigure}{7\textup{a}}
\begin{figure*}

\includegraphics{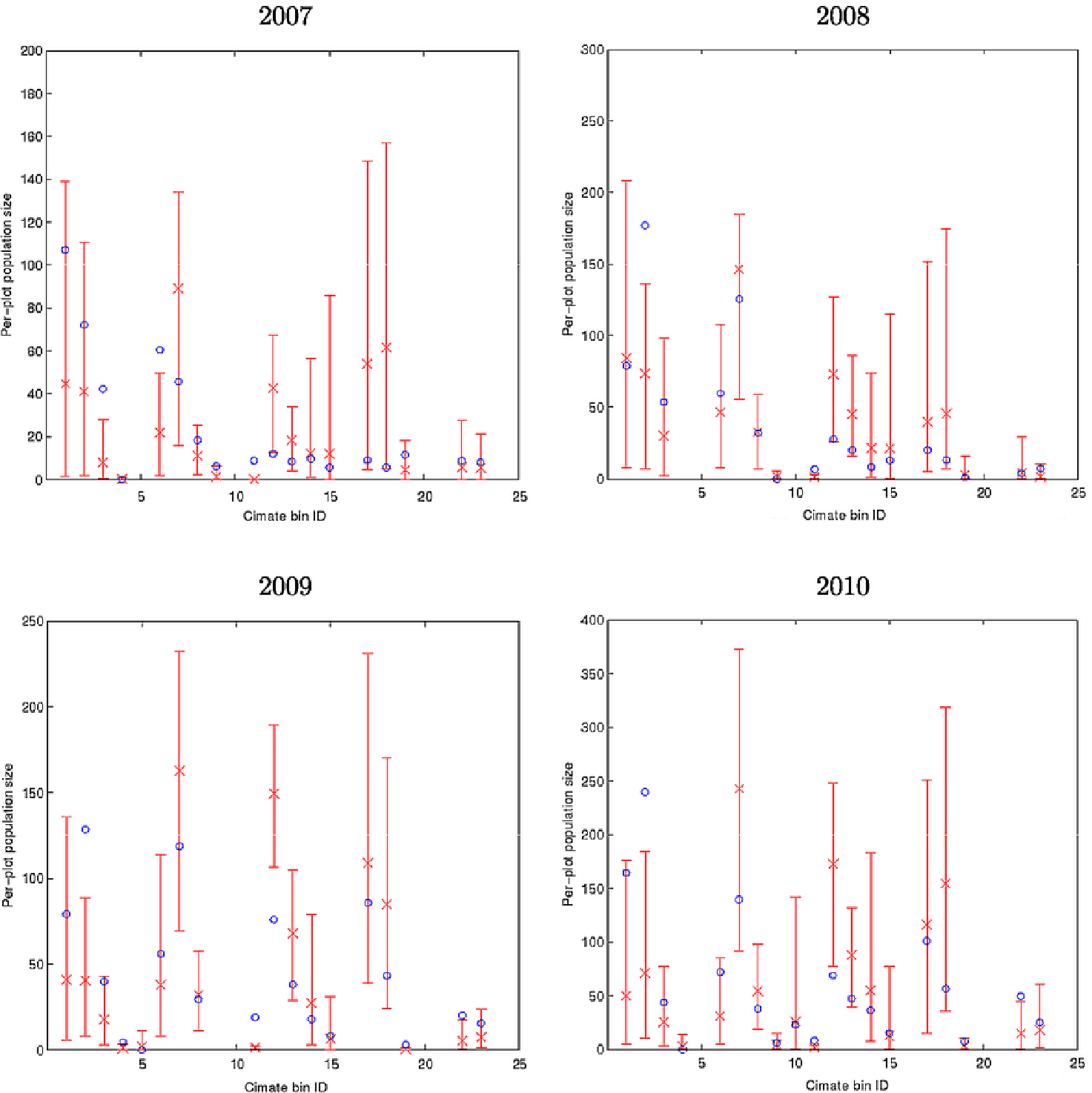}

\caption{Posterior mean (x) and 95\% CI for the estimated
per-plot abundance for ACRU for nonempty climate bins obtained for
years 2007 through 2010. The observed per-plot abundance for the
corresponding climate bins are overlaid (o). The bins are indexed on
the $x$-axis following Figure \protect\ref{fig4}.}\label{fig7a}
\end{figure*}

%
%
\renewcommand{\thefigure}{7\textup{b}}
\begin{figure*}

\includegraphics{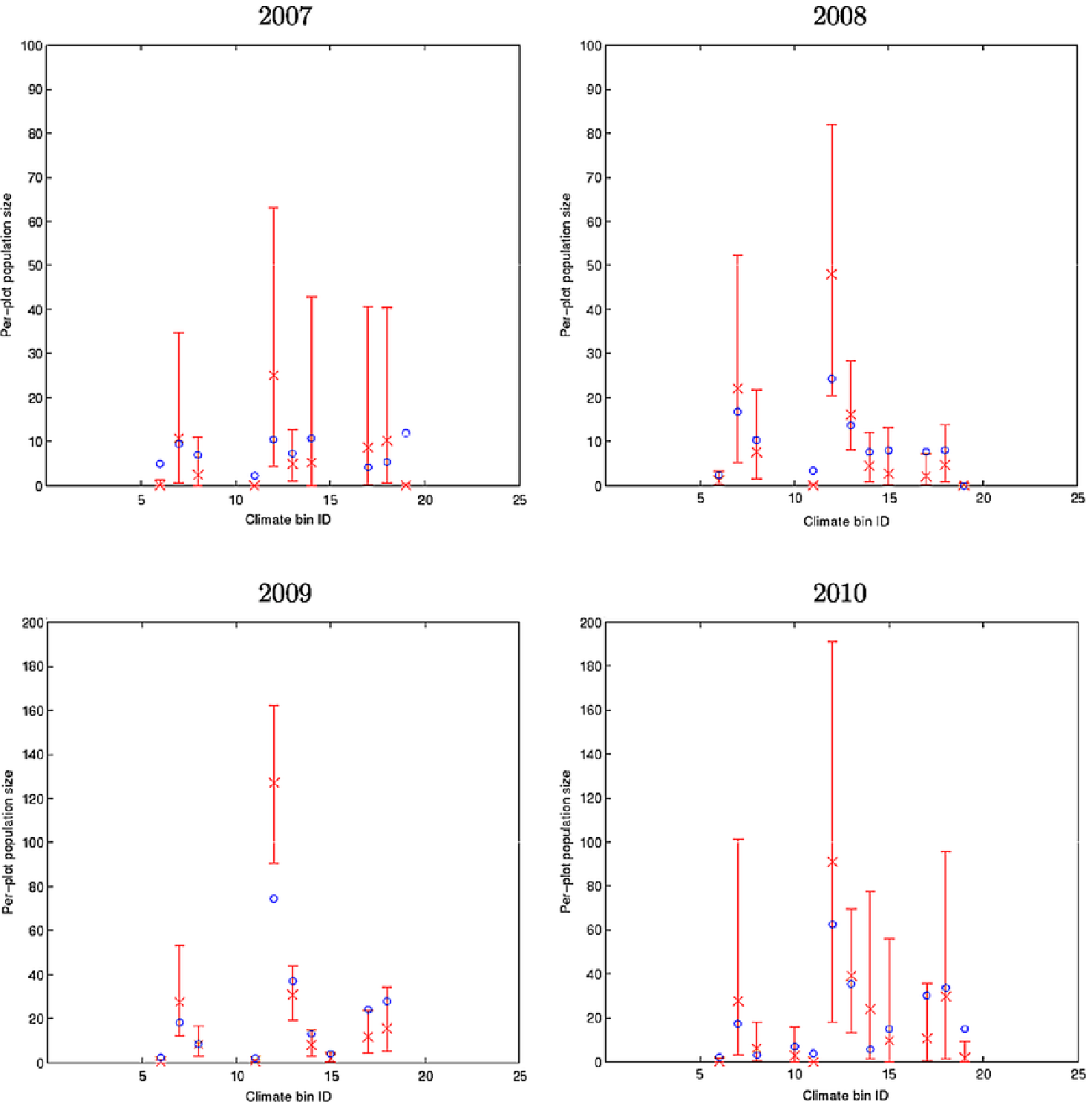}

\caption{Posterior mean ($x$) and 95\% CI for the estimated
per-plot abundance for LITU for nonempty climate bins obtained for
years 2007 through 2010. The observed per-plot abundance for the
corresponding climate bins are overlaid (o). The bins are indexed on
the $x$-axis following Figure
\protect\ref{fig4}.}\label{fig7b}
\end{figure*}

The posterior summaries of the parameters for both species are shown in
Table~\ref{tabl2}. We assumed the upper bound for survival
probability for both ACRU and LITU to be 0.9 and the estimates of $Q_1$
suggest a stronger density dependence for ACRU as
compared to LITU. The recruitment rate ($\Delta$) for ACRU is higher
than that for LITU, explaining the relatively higher
abundance of the former compared to the latter. The climate covariates
do not seem to significantly impact the evolution of
population size for either species. We attribute this, again, to the
severe sparsity which leads to high uncertainty in these
parameters, manifested by very wide credible intervals. Projection
would proceed at the plot level, assuming we know climate in
the intervening, unobserved years.


\section{A Brief Summary and Future Work}\label{sec6}

Only in recent work of \citet{GhoGelCla12} have IPM models been
considered at the population scale and never have IPM models been
considered at large regional scales or in the absence of data for
consecutive time periods. Here we have presented a modeling
approach to enable this in the context of an important demographic data
set, the FIA data which samples plots, not individuals,
roughly every five years, for the entire eastern half of the United
States. After specifying our population level IPM, we have
shown how to scale this IPM from plots to large regions and then we
have shown how to approximately fit this scaled specification
in the presence of the more than $80\%$ absence in the IPM data. We
have illustrated the analysis for two species in the FIA
data set.

Future work will see us investigating additional species in the FIA
database (there are roughly 100 and many are not prevalent)
to compare IPMs. We will also explore the possibility of building a
joint IPM specification to allow for dependence between
species. Such dependence could obviously affect both population size
and diameter distribution. Scaling such joint models from
plots to large geographic regions will add further challenge.

%
\begin{appendix}\label{app}
\section*{Appendix}

The redistribution kernel $K$ is specified as a parametric form. We
assume $K$ to be comprised of growth and recruitment with
climate scaling and density dependence in the form
%
%
\renewcommand{\theequation}{\arabic{equation}}
\begin{eqnarray}\label{genK}\label{equ10}\qquad
K_t(y,x;\mathbf{z}_t,\btheta,\gamma_{t,\cdot})
&=&
\bigl(G_t(y,x;\mathbf{z}_t,\btheta,
\gamma_{t,\cdot})\nonumber\\[-8pt]\\[-8pt]
&&{}+R_t(y;\mathbf{z}_t,\btheta,\gamma
_{t,\cdot})\bigr)e^{\mathbf{z}_{t}^{T}\bfbeta}.\nonumber
\end{eqnarray}
In (\ref{genK}), the exponential term implies multiplicative scaling
of climate effects regardless of $x$. It is introduced
illustratively and to facilitate model identifiability and fitting;
thus, the $\mathbf{z}_{t}$'s are removed from the $G$ and $R$
terms. We have considered other forms for $K$. For instance, climate
might drive growth, that is, be introduced in $f$ while
population size or growth could be introduced\vadjust{\goodbreak} to drive survival. The
flexibility in specifying $K$ is attractive, but the more
complex $K$ is, the weaker the identifiability, the greater the
sensitivity to prior specification, the more difficult the model
fitting. The suggestion is that simple forms for the vital rates below,
which determine $K$, may be more sensible.

In fact, the growth term $G_t(\cdot)$ is further decomposed as
\[
G_t(y|x; \btheta,\gamma_{t,\cdot})=q(x, \gamma_{t,\cdot})
f_{t}(y-x; \btheta).
\]
%
Again, returning to density elements, we interpret $q(x, \gamma_{t,\cdot})
f_{t}(y-x; \btheta) \gamma_{t}(x) \,dx \,dy$ as the expected
number of individuals in diameter interval $(y, y+dy)$ at time $t+1$
from survivors in diameter interval $(x,x+dx)$ at time $t$.
In particular, we assume $f_{t}$ to be Gaussian density. Note that a
translation-invariant $f_{t}$ can be appropriate at the
population level, though it would almost never be sensible at the
individual level. In the sequel, again for convenience, we
assume survival probability declines as a function of $\gamma_{t,\cdot}$
due to resource limitation (but not as a function of diameter)
and consider a logit form,
%
%
\begin{equation}\label{qt}\label{equ11}
q(\gamma_{t,\cdot})=\frac{Q_0 e^{-Q_1 \gamma_{t,\cdot}}}{1+Q_0
e^{-Q_1\gamma
_{t,\cdot}}},
\end{equation}
where $Q_0$ and $Q_1$ (both $>0$) are parameters that govern the rate
of decay of the survival probability.

The recruitment term takes a form similar to the growth term,
\[
R_t(y; \btheta,\gamma_{t,\cdot})=\Delta(x,
\gamma_{t,\cdot}) g_{t}(y; \btheta).
\]
With density elements, analogously, we interpret\break  $\Delta(x,\gamma
_{t,\cdot}) g_{t}(y; \btheta) \gamma_{t}(x) \,dx \,dy$ as the expected
number of recruited individuals in diameter interval $(y, y+dy)$ at
time $t+1$ from individuals in diameter interval $(x, x+dx)$
at time $t$. Usually, the terms on the right-hand side reflect
flowering and
seed production (see, e.g., \citecs{ReeEll09}).
However, with trees, as in our FIA data set, seeds are not monitored.
Hence, the recruitment simply describes the diameter
intensity for new trees in year $t+1$. $\Delta$~is the expected influx
in year $t+1$ and $g_{t}$ is a diameter density on
$\tilde{y}=y-L$ (since all new recruits to our point patterns are at
least size $L$ in the year they arrive), which is assumed to
be an exponential distribution translated to $[L, \infty)$. We assume
influx declines with $\gamma_{t,\cdot}$ due to reduced\vadjust{\goodbreak}
availability of resources and consider the form
%
%
\begin{equation}\label{deltat}\label{equ12}
\operatorname{log}\Delta(\gamma_{t,\cdot})=\delta_0 -
\delta_{1} \gamma_{t,\cdot}
\end{equation}
with $\delta_0$ and $\delta_1$ both nonnegative.

With the resulting kernel inserted into (\ref{equ10}), integrating over $y$, we obtain
%
%
\begin{equation}\label{Gammat+1}\label{equ13}
\gamma_{t+1,\cdot}=\bigl(q(\gamma_{t,\cdot})+\Delta(
\gamma_{t,\cdot})\bigr)e^{\mathbf
{z}_{t}^{T}\bfbeta} \times\gamma_{t,\cdot},
\end{equation}
which clarifies how the expected number of individuals changes from
time $t$ to time $t+1$.
Evidently, we can propagate (\ref{equ13}) across $t$ to learn about the behavior
of population size over time.

Returning to (\ref{equ3}), we approximate the stochastic integral with a
Riemann\vadjust{\goodbreak}
sum. We divide the interval $[L,U]$ into a fine grid
consisting of $B$ cells of equal length with the centers given by $x^*_j$.
We assume that the intensity is constant within each cell and that the
centers, $x^*_j$, remain fixed across all of the time
periods. The length and cell level intensity for cell $b$ are denoted
by $d$ and $\lambda_t(b); b=1,\ldots,B$, respectively. Then
the operational likelihood becomes\looseness=1
%
%
\begin{equation}\label{likelihoodnew}\label{equ14}
\prod_{t=1}^{T} \Biggl[ \exp\Biggl( -\sum
_{b=1}^B \lambda_t(b)d \Biggr)
\prod_{b=1}^{B}\bigl[\lambda_t(b)
\bigr]^{n_{tb}} \Biggr],
\end{equation}\looseness=0
where $n_{tb}$ is the number of points in cell $b$ in year $t$.%

As noted above, we assume $f_{t}(y-x;\mu_t,\sigma_{t}^{2})=\phi
(y-x;\mu_t,\sigma_{t}^{2})$ and $g_{t}(y;\eta_t)=\eta_t
e^{-\eta_t(y-L)}, y>L$.
For the forms in (\ref{equ11}) and (\ref{equ12}), imposing priors on $q(\gamma_{t,\cdot})$
and $\Delta(\gamma_{t,\cdot})$ requires specifying priors on
%
%
\renewcommand{\thetable}{2}
\begin{table*}
\caption{Posterior summary of model parameters for ACRU and for
LITU. Posterior mean and $95\%$ equal tail credible intervals~provided}
\label{tabl2}
\begin{tabular*}{\tablewidth}{@{\extracolsep{\fill}}l l l@{}}
\hline
\multicolumn{1}{@{}l}{\textbf{Parameters}} &
\multicolumn{1}{c}{\textbf{Posterior summary for ACRU}}
& \multicolumn{1}{c@{}}{\textbf{Posterior summary for LITU}} \\
\hline
$Q_{1}$ & 0.08 (0.009, 0.19) & \hphantom{$-$}0.06 (0.0052, 0.099) \\
$\sigma_{}$ & 0.38 (0.08, 0.49) & \hphantom{$-$}0.45 (0.28, 0.72)\\
$\delta_{}$ & 0.19 (0.002, 0.47) & \hphantom{$-$}0.16 (0.001, 0.47) \\
$\eta_{}$ & 0.08 (0.018, 0.14) & \hphantom{$-$}0.07 (0.017, 0.15) \\
$\beta_{\mathrm{intercept}}$ & 0.0746 ($-$5.16, 5.02) & \hphantom{$-$}1.96 ($-$5.90, 10.06)\\
$\beta_{\mathrm{temp}}$ & 0.02 ($-$0.28, 0.30) & $-$0.16 ($-$0.59, 0.24) \\
$\beta_{\mathrm{precip}}$ & 0.0023 ($-$0.0019, 0.0059) & \hphantom{$-$}0.0014 ($-$0.0043,
0.0061) \\
\hline
\end{tabular*}
\end{table*}
$Q_0, Q_1, \delta_0$ and $\delta_1$, respectively. We interpret
$\frac{Q_0}{1+Q_0}$ as the survival probability when the
population size tends to $0$ and $\frac{\delta_0}{1+\delta_0}$ as
the replacement rate when the population size tends to $0$. We
can roughly interpret $Q_1$ to be the global survival probability of
the species and $\delta_1$ to be the average rate of
influx shown by that species.

$Q_0, Q_1, \delta_0$ and $\delta_1$ are not well identified. In fact,
from (\ref{equ13}), the sum $q(\cdot) + \Delta(\cdot)$ is well identified
but not its components. Estimation of $q(\gamma_{t,\cdot})$ and $\Delta
(\gamma_{t,\cdot})$ requires using knowledge of the ecological
processes driving the survival and influx for the population. According
to the species, we assume a known upper bound of the
survival and recruitment function which are achieved when $\gamma
_{t,\cdot}=0$. Solving these\break boundary conditions, we obtain the
values of $Q_0$ and $\delta_0$ and do not estimate them as part of
model fitting. $Q_1$ and $\delta_1$, on the other hand, are
estimated as a part of fitting using an additional constraint. Let
$\rho_{t+1,t}=(N_{t+1}-N_t)/N_t, t=0,1,\ldots,T-1$, where $N_t$
is the total observed population size in year $t$. Then $\rho_t$
denotes the relative change in population size in two
consecutive years. To induce identifiability, we impose that $q(\gamma
_{t,\cdot})+\Delta(\gamma_{t,\cdot}) \in(1-\max_t
(\rho_{t+1,t}),1+\max_t(\rho_{t+1,t}))$. The priors on $Q_1$ and
$\delta_1$ are chosen such that this constraint is satisfied.

The $\bfbeta$'s are well identified since they are regression
coefficients associated with time varying covariates $\mathbf{z}_t$.
Hence, we impose a vague $\operatorname{Normal}(0,100)$ prior on each
component of
$\bfbeta$ independently.
\end{appendix}

\section*{Acknowledgments}

The authors acknowledge useful conversations with Kai Zhu (who also
provided assistance with the data preparation), Thomas M{\o}lhave (who
also facilitated the climate space partitioning) and Pankaj Agarwal in
building upon earlier work to develop this manuscript. The work of the
authors was supported in part by NSF DEB 0516320, NSF DMS-09-14906 and
NSF CDI 0940671.


%

\end{document}